\documentclass[12pt,prd,tightenlines,nofootinbib]{revtex4}

\usepackage{bm}
\usepackage{graphicx}
\usepackage{multirow}
\begin{document}
\title{\bf Heavy tetraquarks in the relativistic quark model }

\author{Rudolf N. Faustov$^{1}$, Vladimir
  O. Galkin$^{1}$ and Elena M. Savchenko$^{2}$}

\affiliation{%
$^{1}$  Institute of Cybernetics and Informatics in Education,  Federal
Research Center ``Computer Science and Control'', Russian Academy of
Sciences, Vavilov Street 40, 119333 Moscow\\
$^{2}$  Faculty of Physics, M.V. Lomonosov Moscow State
University,  119991 Moscow, Russia}

\begin{abstract}
{We give a review of the calculations of the masses of
  tetraquarks with two and four heavy quarks in the framework of the relativistic quark model based on the quasipotential
  approach and QCD. The diquark-antidiquark picture of heavy tetraquarks is
  used. The quasipotentials of the quark-quark and
diquark-antidiquark interactions are constructed similarly to the
previous consideration of mesons and  baryons. Diquarks are considered
in the colour triplet state. It is assumed that the diquark and
antidiquark interact in the tetraquark as a whole and the internal
structure of the diquarks is taken into account by the calculated form factor of the diquark-gluon interaction. All parameters of the
model are kept fixed from our previous calculations of meson and
baryon properties. A detailed comparison of the obtained predictions
for heavy tetraquark masses with available experimental data is
given. Many candidates for tetraquarks are found. It is argued that the
structures in the di-$J/\psi$ mass spectrum observed recently by the
LHCb Collaboration can be interpreted as $cc\bar c\bar c$
tetraquarks. }
\end{abstract}
\maketitle

\section{Introduction}

The possibility of the existence of exotic multiquark hadrons with the
content of the valence
quarks and antiquarks different from a quark-antiquark pair for mesons and
three quarks for baryons, had been considered since the early days of the
quark model. However, the absence of the convincing experimental evidence
for such multiquark states  made their investigation of marginal
interest for several decades. The situation dramatically changed in the
last two decades.   This subject
became a hot topic since the first explicit experimental evidence of the
existence of hadrons with compositions different from usual $q\bar q$
for mesons and $qqq$ for baryons became available (for recent reviews, see \cite{Liu:2019zoy,Brambilla:2019esw,ghmwzz,osz,betal,lms} and
references therein).  Candidates for both the exotic tetraquark
$qq\bar q\bar q$ and pentaquark
$qqqq\bar q$ states were found. However, in the literature there is no
consensus about the composition of these states \cite{Liu:2019zoy,Brambilla:2019esw,ghmwzz,osz,betal,lms}. For example,
significantly different interpretations for the
$qq\bar q\bar q$ candidates were proposed: molecules
composed from two mesons loosely bound by the meson exchange,
compact  tetraquarks composed from a diquark and antidiquark bound by
strong forces,
hadroquarkonia composed of a heavy quarkonium embedded in a light
meson, kinematic cusps, etc. The discrimination between different
approaches is a very complicated experimental task.

The simplest multiquark system is a tetraquark, composed of two
quarks and two antiquarks. Heavy tetraquarks are of particular
interest, since the presence of a heavy quark increases the
binding energy of the bound system and, as a result, the
possibility that such tetraquarks will have masses below the
thresholds for decays to mesons with open heavy flavour. In this case the
strong decays, which proceed through the quark and antiquark
rearrangements, are kinematically forbidden, and the corresponding
tetraquarks can decay only weakly or electromagnetically and thus
they should have a tiny decay width.  If the predicted tetraquarks
have masses slightly (a few MeV) above these thresholds, then they can
be also observed as resonances. The excited tetraquark states could be
also narrow, notwithstanding the large phase space, since their decays
will be suppressed either by the centrifugal barrier between quarks and
antiquarks or by the nodes of the wave function of
radially excited states, or even both.

\begin{table}
\caption{Experimental data on hidden-charm exotic
  mesons}\label{ccccexp}
\begin{ruledtabular}
\begin{tabular}{@{}c@{ } c@{ } c@{ } c@{ }c@{ }c@{}}
  State            & $J^{PC}$          & $M$ (MeV) & $\Gamma$ (MeV) &
                                                                      Observed in & Experiment \\ \hline
  $X(3872)$        & $1^{++}$      & $3871.69\pm0.17$         & $<1.2$         & $B^\pm\to K^\pm \pi^+ \pi^- J/\psi$  & Belle \\
  $Z_{c}(3900)^\pm$    & $1^{+-}$     & $3888.4\pm2.5$         &
                                                                 $28.3\pm2.5$   & $e^+ e^- \to \pi^+ \pi^- J/\psi$   & BESIII \\
  $X(3940)$&$?^{??}$   &$3942\pm9$&$37^{+27}_{-17}$&$e^+e^-\to J/\psi
                                                     X$& Belle\\
  $Z_{cs}(3985)^-$&$1^{+}$  &$3982.5^{+1.8}_{-2.6}\pm2.1$&$12.8^{+5.3}_{-4.4}\pm3.0$&$e^+e^-\to K^+(D_s^-D^{*0}+D^*_sD^0)$&BESIII\\
  $Z_c(4020)^\pm$  & $?^{?-}$   & $4024.1\pm1.9$             & 
                                                               $13\pm5$      & $e^+ e^- \to \pi^+ \pi^- h_c(1P)$ & BESIII \\
$Z_c(4050)^\pm$ & $?^{?+}$ & $4051\pm14^{+20}_{-41}$ & $82^{+21+47}_{-17-22}$&  
                                                                   $\bar B^0 \to K^- \pi^+ \chi_{c1}(1P)$  
                                                                                                       &Belle\\
$Z_c(4055)^\pm$ & $?^{?-}$ & $4054\pm3\pm1$ & $45\pm11\pm6$& $e^+ e^- \to 
                                                      \psi(2S)\pi^+\pi^-$
                                                                                                       &Belle\\
$Z_c(4100)^\pm$ & $?^{??}$ & $4096\pm20^{+18}_{-22}$ & $152\pm58^{+60}_{-35}$& $B^0 \to K^+ \pi^- \eta_c$ & LHCb \\       
  $X(4140)$        & $1^{++}$      & $4146.8\pm2.4$           & 
                                                                $22^{+8}_{-7}$ & $ B^{+} \to \phi J/\psi K^{+}$   & CDF, LHCb \\ 
  $Z_c(4200)^\pm$  & $1^{+-}$    & $4196^{+31+17}_{-29-13}$ & 
                                                             $370\pm70^{+70}_{-132}$ &  $\bar B^0 \to K^- \pi^+ J/\psi$   
                                                                                                       &Belle\\
  $Y(4230)$        & $1^{--}$        & $4218.7\pm2.8$               & 
                                                                  $44\pm9$ 
                                                                                   &  $e^+ e^- \to \omega\chi_{c0}$& BESIII\\             
  $Z_c(4240)^\pm$   & $0^{--}$    & $4239\pm18^{+45}_{-10}$ & 
                                                             $220\pm47^{+108}_{-74}$ & $B^0 \to K^+ \pi^- \psi(2S)$ & LHCb \\ 
$Z_c(4250)^\pm$ & $?^{?+}$ & $4248^{+44+180}_{-29-35}$ & $177^{+54+316}_{-39-61}$&   
                                                                   $\bar B^0 \to K^- \pi^+ \chi_{c1}(1P)$   
                                                                                                       &Belle\\                  
  $Y(4260)$        & $1^{--}$        & $4230\pm8$               &
                                                                  $55\pm19$      & $e^+ e^- \to \gamma_{\rm ISR}\pi^+ \pi^- J/\psi$& BaBar \\
  $X(4274)$        & $1^{++}$      & $4274^{+8}_{-6}$     & $49\pm 12$ & $B^+ \to J/\psi \phi K^+$  & CDF, LHCb \\
  $Y(4360)$        & $1^{--}$        & $4368\pm13$               &
                                                                  $96\pm7$
                                                                                   &
                                                                                     $e^+ e^- \to \gamma_{\rm ISR}\pi^+ \pi^- \psi(2S)$  & Belle  \\
  $Y(4390)$        & $1^{--}$        & $4392\pm7$               & 
                                                                  $140^{+16}_{-21}$  & $e^+ e^- \to \pi^+ \pi^- h_c$ & BESIII \\  
  $Z_c(4430)^\pm$  & $1^{+-}$   & $4478^{+15}_{-18}$       & $181\pm31$     & $B \to K \pi^\pm \psi(2S)$   & Belle \\
  $X(4500)$        & $0^{++}$      & $4506\pm11^{+12}_{-15}$       & $92\pm21^{+21}_{-20}$      & $B^+ \to J/\psi \phi K^+$   & LHCb \\ 
  $Y(4630)$        & $1^{--}$        & $4634^{+8+5}_{-7-8}$         & $92^{+40+10}_{-24-21}$& $e^+ e^- \to \Lambda^{+}_{c} \Lambda^{-}_{c}$ & Belle \\ 
  $Y(4660)$        & $1^{--}$        & $4633\pm7$               & 
                                                                  $64\pm9$
                                                                                   & $e^+ e^- \to \gamma_{\rm ISR}\pi^+ \pi^- \psi(2S)$   & Belle \\
  $X(4700)$        & $0^{++}$      & $4704\pm10^{+14}_{-24}$       & 
                                                                $120\pm31^{+42}_{-33}$     & $B^+ \to J/\psi \phi K^+$   & LHCb \\ 
 $X(4740)$        & $?^{?+}$     & $4741\pm 6 \pm 6$       &  
                                                                $53 \pm 15 \pm 11$ 
                                                                    &
                                                                      $B_s \to J/\psi \phi \pi^+\pi^-$   & LHCb \\
  $X(6900)$&$?^{?+}$&$6905\pm11\pm7$&$80\pm19\pm33$&$pp\to J/\psi
                                                   J/\psi X$& LHCb\\
  \end{tabular}
\end{ruledtabular}
\end{table}

In Table~\ref{ccccexp} we collect experimental data on hidden-charm
mesons with exotic properties \cite{pdg,Zcs,Y4630,Aaij:2020fnh,Aaij:2020tzn}. We use the $XYZ$ naming scheme, where
$X$ are neutral exotic charmonium-like states, observed in hadronic decays, $Y$ are neutral exotic
charmonium-like states with $J^{PC}=1^{--}$, observed in $e^+e^-$
collisions, and $Z$ are charged (isospin triplet $I=1$)
charmonium-like states. The later ones are explicitly exotic, since
they could not be simply $c\bar c$ states and in order  to
have a nonzero charge these states should contain at least
additional light quark and antiquark.  The experimentally determined quantum
numbers $J^{PC}$, masses $M$, total decay widths $\Gamma$, observation
channels and names of the experiments where they were first observed
are given  in Table~\ref{ccccexp} \cite{pdg,Aaij:2020fnh,Aaij:2020tzn}. To determine the quantum numbers of $X$ and $Z$ states a
rather complicated angular analysis was necessary, while those of $Y$
states, which coincided with the quantum numbers of the photon, are
determined by the observation channel. Note that the $X(4140)$, $X(4274)$,
$X(4500)$, $X(4700)$ and $X(4740)$ states were observed as resonances in the
$J/\psi\phi$ mass spectrum, thus they should contain the strange quark and strange
antiquark instead of the $u$ and $d$ quarks and antiquarks.
Very recently the BESIII Collaboration \cite{Zcs} reported the first
candidate for the charged charmonium-like state with the open
strangeness $Z_{cs}(3885)$.  
It is
important to point out that most of these exotic states have masses 
close to the thresholds of the open and/or hidden flavor meson
production.

Many theoretical interpretations of these states were
suggested in the literature (for recent reviews, see
\cite{Liu:2019zoy,Brambilla:2019esw,ghmwzz,osz,betal,lms} and references
therein). Main of them are the following. 
The conventional $c\bar c$ states influenced by
the open flavor thresholds. It is clear that such interpretation is
inapplicable at least to the charged $Z$ states. However, the $c\bar
c$ admixture may be present in some of the neutral states. Thus exotic interpretations
were proposed. They include:
\begin{enumerate}
  \item Molecules, which are two heavy
 mesons ($Q\bar q)(\bar Q q)$, loosely bound by meson exchanges \cite{ghmwzz}; 
\item {Tetraquarks, which are 
 four-quark $Q  q\bar Q\bar q$ states, tightly bound by the color forces}~\cite{i,mppr,efgtetr};
 \item Hybrids, which are $Q\bar Q$-gluon states with
excited gluonic degrees of freedom~\cite{os};
\item Hadro-quarkonium, which are  compact quarkonium states $Q\bar Q$
  embedded in an excited light-quark matter~\cite{v};
\item Kinematic or rescattering effects at corresponding thresholds~\cite{vbr};
 { \item  $Q\bar Q$ core plus molecular-like components~\cite{vbr1}.} 
\end{enumerate}

 In this review we consider these exotic heavy mesons as heavy
 tetraquarks \cite{mppr,efgtetr,efgextetr,efgl,efgb}. Our main
 assumption is the following one. Tetraquarks are composed from a
diquark and antidiquark in color $\bar 3$ and 3 configurations, which are bound
by color forces. This assumption reduces the very complicated
four-body relativistic calculation  to a more simple two-step two-body calculation. First,
a diquark $d$ (antidiquark $\bar d$) is considered as a
$qq'$ ($\bar q \bar q'$) bound state (as in baryons). Note, that only the
color triplet configuration contributes since there is a repulsion
between quarks in a color sextet.
Second, a tetraquark is considered as the
$d\bar d'$ bound state where constituents are assumed to interact as a
whole.  This means that there are no separate interactions between quarks, composing a diquark, and
antiquarks, composing an antidiquark \cite{efgtetr}. The
resulting tetraquark has a typical hadronic size. We consider diquarks
in the ground state only, as in the case of heavy baryons \cite{hbar}. All
excitations are assumed to be in $d\bar d$ bound system.
A rich spectroscopy is predicted since both radial and
orbital excitations can occur between diquarks. However, the number of
predicted excited states is significantly less than in a pure
four-body picture of a tetraquark.

When one constructs a diquark it is necessary to remember that it is a
composite $(qq')$ system. Thus, a diquark is not a point-like
object. Indeed,   its interaction with gluons is smeared by the form
factor which can be expressed through the overlap integral of diquark wave
functions. Also the Pauli principle should be taken into account.  For the
ground state diquarks it leads to the following restrictions.  The  $(qq')$
diquark, composed from quarks of different flavours,  can have spins
$S=0,1$ (scalar $[q,q']$, axial vector $\{q,q'\}$ diquarks, while 
 the $(qq)$ diquarks, composed from quarks of the same flavour, can
 have only $S=1$ (axial vector $\{q,q\}$ diquark). The scalar $S$
 diquark is more tightly bound and have a smaller mass because of  the larger attraction due
 to the spin-spin interaction. It is often called a ``good'' diquark and
the heavier axial vector $A$ diquark is called a ``bad'' diquark.   

It is important to emphasize that we treat both light and heavy
quarks and diquarks fully relativistically without application of the nonrelativistic ($v/c$) expansion.

   In this review we consider the following tetraquarks.
\begin{enumerate}
\item Heavy tetraquarks  $(Qq)(\bar Q\bar
  q')$ with hidden charm and bottom \cite{efgtetr,efgextetr,efgb}. \\
  The neutral $X$ should be split into two states  ($[Qu][\bar Q\bar u]$ and
$[Qd][\bar Q\bar d]$) with $\Delta M\sim $ few~MeV.
The model predicts the existence of their charged partners  $X^+=[Qu][\bar Q\bar d]$,
$X^-=[Qd][\bar Q\bar u]$ and the existence of tetraquarks with open $X_{s\bar q}=[Qs][\bar
Q\bar q]$ and hidden  $X_{s\bar s}=[Qs][\bar Q\bar s]$ strangeness.
\item Doubly heavy tetraquarks  $(QQ')(\bar q\bar
  q')$ with open charm and bottom \cite{efgl}.\\ 
These tetraquarks are explicitly exotic with
heavy flavor number equal to 2. 
Their observation would be a direct
proof of the existence of multiquark states. The estimates of the
production rates of such tetraquarks indicate that they could be
produced and detected at present  and future facilities.  We considered the doubly heavy
$(QQ')(\bar q\bar q')$ tetraquark ($Q,Q'=b,c$ and $q,q'=u,d,s$) as the bound
system of the heavy 
diquark ($QQ'$) and light antidiquark ($\bar q\bar q'$). 
\item Heavy tetraquarks  $(cq)(\bar b\bar
  q')$ with open charm and bottom \cite{efgl}.\\
We considered heavy
$(cq)(\bar b\bar q')$ tetraquark ($q,q'=u,d,s$) as the bound
system of the heavy-light 
diquark ($cq$) and heavy-light antidiquark ($\bar b\bar q'$). 
\item $QQ\bar Q\bar Q$ tetraquarks composed from
  heavy ($Q=c,b$) quarks only \cite{fgs}.\\
The new structures in
double-$J/\psi$ spectrum have been very recently observed by the LHCb
Collaboration in proton-proton collisions \cite{Aaij:2020fnh}. 
On the other hand, the absence of narrow
structures in the $\Upsilon$-pair production was reported by the LHCb \cite{Aaij:2018zrb} and CMS \cite{Sirunyan:2020txn} Collaborations.
We considered heavy
$(QQ')(\bar Q\bar Q')$ tetraquark as the bound system of the doubly
heavy diquark ($QQ'$) and doubly heavy antidiquark ($\bar Q\bar Q'$).
 \end{enumerate}
   
\section{Relativistic diquark-antidiquark model of heavy tetraquarks}
\label{sec:rqm}

For the calculation of the masses of tetraquarks we use the
relativistic quark model based on the quasipotential approach and the
diquark-antidiquark picture of tetraquarks. First, we calculate the
masses  and  wave functions ($\Psi_{d}$) of the light and heavy diquarks as the bound quark-quark states.  Second,  the masses of the tetraquarks  and their wave functions ($\Psi_{T}$) are obtained for the bound diquark-antidiquark states. These wave functions are solutions of the  Schr\"odinger-type quasipotential equations \cite{efgtetr,efgl}
\begin{equation}
\label{quas}
{\left(\frac{b^2(M)}{2\mu_{R}}-\frac{{\bf
p}^2}{2\mu_{R}}\right)\Psi_{d,T}({\bf p})} =\int\frac{d^3 q}{(2\pi)^3}
 V({\bf p,q};M)\Psi_{d,T}({\bf q}),
\end{equation}
with the  on-mass-shell relative momentum squared  given by
\begin{equation}
{b^2(M) }
=\frac{[M^2-(m_1+m_2)^2][M^2-(m_1-m_2)^2]}{4M^2},
\end{equation}
and the relativistic reduced mass
\begin{equation}
\mu_{R}=\frac{E_1E_2}{E_1+E_2}=\frac{M^4-(m^2_1-m^2_2)^2}{4M^3}.
\end{equation}
The  on-mass-shell energies  $E_1$, $E_2$ are defined as follows
\begin{equation}
\label{ee}
E_1=\frac{M^2-m_2^2+m_1^2}{2M}, \quad E_2=\frac{M^2-m_1^2+m_2^2}{2M}.
\end{equation}
The bound-state masses of a diquark or a tetraquark are $M=E_1+E_2$ , where
$m_{1,2}$ are the masses of quarks ($Q_1$ and $Q_2$) which form
the diquark or of the diquark ($d$) and antidiquark ($\bar d'$) which
form the heavy tetraquark ($T$), while ${\bf p}$ is their relative
momentum. 

The quasipotential operator $V({\bf p,q};M)$ in Eq.~(\ref{quas}) is  constructed with the help of the off-mass-shell scattering amplitude, projected onto the positive-energy states.
The quark-quark  ($QQ'$) interaction quasipotential \footnote{We
  consider diquarks in a tetraquark, as in a baryon, to be in the color triplet state, since in the
  color sextet there is a repulsion between two quarks.} is considered to be 1/2 of the quark-antiquark   ($Q\bar Q'$) interaction and is given by \cite{hbar}
 \begin{equation}
\label{qpot}
V({\bf p,q};M)=\bar{u}_{1}(p)\bar{u}_{2}(-p){\cal V}({\bf p}, {\bf
q};M)u_{1}(q)u_{2}(-q),
\end{equation}
with
\[
{\cal V}({\bf p,q};M)=\frac12\left[\frac43\alpha_sD_{ \mu\nu}({\bf
k})\gamma_1^{\mu}\gamma_2^{\nu}+ V^V_{\rm conf}({\bf k})
\Gamma_1^{\mu}({\bf k})\Gamma_{2;\mu}(-{\bf k})+
 V^S_{\rm conf}({\bf k})\right].
\]
Here,  $D_{\mu\nu}$ is the
gluon propagator in the Coulomb gauge, $u(p)$ are the Dirac spinors and $\alpha_s$ is the running QCD coupling constant with freezing
\begin{equation}
  \label{eq:alpha}
  \alpha_s(\mu^2)=\frac{4\pi}{\displaystyle\left(11-\frac23n_f\right)
\ln\frac{\mu^2+M_B^2}{\Lambda^2}}, 
\end{equation}
where the scale $\mu$ is chosen to be equal to $2m_1 m_2/(m_1+m_2)$, the background mass is $M_B=2.24\sqrt{A}=0.95$~GeV, and $n_f$ is the number of flavours \cite{alpha}.
 The effective long-range vector vertex  contains both the Dirac and Pauli terms
\cite{efgm} 
\begin{equation}
\Gamma_{\mu}({\bf k})=\gamma_{\mu}+
\frac{i\kappa}{2m}\sigma_{\mu\nu}\tilde k^{\nu}, \qquad \tilde
k=(0,{\bf k}),
\end{equation}
where $\kappa$ is the long-range anomalous chromomagnetic moment. In the nonrelativistic limit the vector and scalar confining potentials in configuration space have the form
\begin{eqnarray}
V^V_{\rm conf}(r)&=&(1-\varepsilon)(Ar+B),\qquad V^S_{\rm conf}(r)=\varepsilon (Ar+B),\nonumber\\[1ex]
V_{\rm conf}(r)&=&V^V_{\rm conf}(r)+V^S_{\rm conf}(r)=Ar+B,
\end{eqnarray}
where $\varepsilon$ is the mixing coefficient.  Therefore in the nonrelativistic limit the $QQ'$ quasipotential reduces to
\begin{equation}
V^{\rm NR}_{QQ'}(r)=\frac12V^{\rm NR}_{Q\bar Q'}(r)=\frac12\left(-\frac43\frac{\alpha_s}{r}+Ar+B\right),
\end{equation} 
reproducing the usual Cornel potential. Thus our quasipotential can be
viewed as its relativistic generalization.  It contains both
spin-independent and spin-dependent relativistic contributions. 

Constructing the diquark-antidiquark ($d\bar d'$) 
quasipotential we use the same assumptions about the structure of the
short- and long-range interactions. Taking into account the integer
spin of a diquark in the color triplet state, the quasipotential is given by \cite{efgtetr,efgl}
\begin{eqnarray}
\label{dpot} 
V({\bf p,q};M)&=&\frac{\langle
d(P)|J_{\mu}|d(Q)\rangle} {2\sqrt{E_dE_d}} \frac43\alpha_sD^{
\mu\nu}({\bf k})\frac{\langle d'(P')|J_{\nu}|d'(Q')\rangle}
{2\sqrt{E_{d'}E_{d'}}}\nonumber\\[1ex]
&&+\psi^*_d(P)\psi^*_{d'}(P')\left[J_{d;\mu}J_{d'}^{\mu} V_{\rm
conf}^V({\bf k})+V^S_{\rm conf}({\bf
k})\right]\psi_d(Q)\psi_{d'}(Q'),
\end{eqnarray}
where $\psi_d(p)$ is the wave function of the diquark, 
\begin{equation}
 \label{eq:ps}
 \psi_d(p)=\left\{\begin{array}{ll}1 &\qquad \text{for a scalar
 diquark}\\[1ex]
\varepsilon_d(p) &\qquad \text{for an axial-vector diquark}
\end{array}\right.
\end{equation}
Here the four-vector
\begin{equation}\label{pv}
\varepsilon_d(p)=\left(\frac{(\bm{\varepsilon}_d\cdot{\bf
p})}{M_d},\bm{\varepsilon}_d+ \frac{(\bm{\varepsilon}_d\cdot{\bf
p}){\bf
 p}}{M_d(E_d(p)+M_d)}\right), \qquad \varepsilon^\mu_d(p) p_\mu=0,
\end{equation}
is the polarization vector of the axial-vector diquark with
momentum ${\bf p}$, $E_d(p)=\sqrt{{\bf p}^2+M_d^2}$, and
$\varepsilon_d(0)=(0,\bm{\varepsilon}_d)$ is the polarization
vector in the diquark rest frame. The effective long-range vector
vertex of the diquark  $J_{d;\mu}$ is given by
\begin{equation}
 \label{eq:jc}
 J_{d;\mu}=\left\{\begin{array}{ll}
 \frac{\displaystyle (P+Q)_\mu}{\displaystyle
 2\sqrt{E_dE_d}}&\qquad \text{ for a scalar diquark,}\\[3ex]
-\; \frac{\displaystyle (P+Q)_\mu}{\displaystyle2\sqrt{E_dE_d}}
 +\frac{\displaystyle i\mu_d}{\displaystyle 2M_d}\Sigma_\mu^\nu
\tilde k_\nu
 &\qquad \text{ for an axial-vector diquark,}\end{array}\right.
\end{equation}
where $\tilde k=(0,{\bf k})$. Here, the antisymmetric tensor
$\Sigma_\mu^\nu$ is defined by
\begin{equation}
 \label{eq:Sig}
 \left(\Sigma_{\rho\sigma}\right)_\mu^\nu=-i(g_{\mu\rho}\delta^\nu_\sigma
 -g_{\mu\sigma}\delta^\nu_\rho),
\end{equation}
and the axial-vector diquark spin ${\bf S}_d$ is given by
$(S_{d;k})_{il}=-i\varepsilon_{kil}$; $\mu_d$ is the total
chromomagnetic moment of the axial-vector diquark. We choose
$\mu_d=0$ to make the long-range chromomagnetic interaction of
diquarks, which is proportional to $\mu_d$,  vanish in accordance
with the flux-tube model.
The vertex of the diquark-gluon
interaction $\langle d(P)|J_{\mu}|d(Q)\rangle$  accounts for the internal structure of  
the diquark
\begin{equation}
 \langle d(P) \vert J_\mu (0) \vert d(Q)\rangle
=\int \frac{d^3p\, d^3q}{(2\pi )^6} \bar \Psi^{d}_P({\bf
p})\Gamma _\mu ({\bf p},{\bf q})\Psi^{d}_Q({\bf q}) , 
\end{equation}
where
$\Gamma_\mu ({\bf p},{\bf q})$ is the two-particle vertex function of
  the  diquark-gluon interaction.
It leads to emergence of the form factor $F(r)$ smearing
the one-gluon exchange potential. This form factor is expressed
through the overlap integral of the diquark wave functions.

All parameters of the model were fixed previously \cite{alpha,efgm,efg} from the consideration of meson and baryon properties. They are the following. The constituent heavy quark masses: $m_b=4.88$ GeV, $m_c=1.55$ GeV. The parameters of the quasipotential: $A=0.18$ GeV$^2$, $B=-0.3$~GeV, $\Lambda=413$~MeV; the mixing coefficient of
vector and scalar confining potentials $\varepsilon=-1$; the universal
Pauli interaction constant $\kappa=-1$. Note that the long-range
chromomagnetic interaction of quarks, which is proportional to
$(1+\kappa)$ vanishes for the chosen value of $\kappa$ in accordance
with the flux-tube model.

\begin{figure}
 \begin{center}\includegraphics[width=10.5 cm]{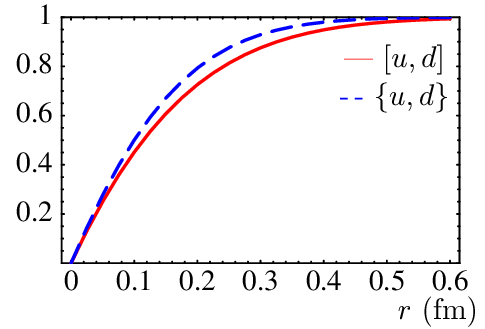}\end{center}
\caption{The form factors $F(r)$ for the scalar
$[u,d]$ (solid line) and axial vector $\{u,d\}$ (dashed line) diquarks.\label{fig:ud}}
\end{figure}

The resulting diquark-antidiquark quasipotential for the tetraquark
states, where quark energies
$\epsilon_{1,2}(p)$ were replaced by the on-shell energies $E_{1,2}$
(\ref{ee}) to remove the non-locality, is given by  \cite{efgl}
\begin{eqnarray}
  \label{eq:pot}
  V(r)&=&V_{\rm Coul}(r)+V_{\rm conf}(r)+\frac12\Biggl\{\left[
   \frac1{E_1(E_1+M_1)}+\frac1{E_2(E_2+M_2)}\right]
\frac{\hat V'_{\rm Coul}(r)}r \cr
&&
-\Biggl[\frac1{M_1(E_1+M_1)} +\frac1{M_2(E_2+M_2)}\Biggr]
\frac{V'_{\rm conf}(r)}r\cr
&& +\frac{\mu_d}2
\left(\frac1{M_1^2}+\frac1{M_2^2}\right)\frac{V'^V_{\rm conf}(r)}r\Biggr\}
{\bf L}\cdot ({\bf
S}_1+{\bf S}_2 )+\frac12\Biggl\{\Bigl[
   \frac1{E_1(E_1+M_1)}\cr
&&-\frac1{E_2(E_2+M_2)}\Bigr]
\frac{\hat V'_{\rm Coul}(r)}r 
-\left[\frac1{M_1(E_1+M_1)}-\frac1{M_2(E_2+M_2)}\right]
\frac{V'_{\rm conf}(r)}r \cr
&& +\frac{\mu_d}2
\left(\frac1{M_1^2}-\frac1{M_2^2}\right)\frac{V'^V_{\rm conf}(r)}r\Biggl\}
{\bf L}\cdot ({\bf
S}_1-{\bf S}_2 )+\frac1{E_1E_2}\Biggl\{{\bf
    p}\left[V_{\rm Coul}(r)+V^V_{\rm conf}(r)\right]{\bf p}\cr
&& -\frac14 
\Delta V^V_{\rm conf}(r)+ V'_{\rm Coul}(r)\frac{{\bf
    L}^2}{2r}
+\frac1{r}\left[V'_{\rm Coul}(r)+\frac{\mu_d}4\left(\frac{E_1}{M_1}
+\frac{E_2}{M_2}\right)V'^V_{\rm conf}(r)\right]{\bf L}
({\bf S}_1+{\bf S}_2)\cr
&&
+\frac{\mu_d}4\left(\frac{E_1}{M_1}
-\frac{E_2}{M_2}\right)\frac{V'^V_{\rm conf}(r)}{r}{\bf L}
({\bf S}_1-{\bf S}_2)
+\frac13\Biggl[\frac1{r}{V'_{\rm Coul}(r)}-V''_{\rm Coul}(r)\cr
&&
+\frac{\mu_d^2}4\frac{E_1E_2}{M_1M_2}
\left(\frac1{r}{V'^V_{\rm conf}(r)}-V''^V_{\rm
    conf}(r)\right)\Biggr]\left[\frac3{r^2}({\bf S}_1{\bf r}) ({\bf
  S}_2{\bf r})- 
{\bf S}_1{\bf S}_2\right]\cr
&&
+\frac23\left[\Delta V_{\rm Coul}(r)+\frac{\mu_d^2}4\frac{E_1E_2}{M_1M_2}
\Delta V^V_{\rm conf}(r)\right]{\bf S}_1{\bf S}_2\Biggr\}.
\end{eqnarray}
Here $$\hat V_{\rm Coul}(r)=-\frac{4}{3}\alpha_s
\frac{F_1(r)F_2(r)}{r}$$ is the Coulomb-like one-gluon exchange
potential which takes into account the finite sizes of the diquark
and antidiquark through corresponding form factors $F_{1,2}(r)$. ${\bf
  S}_{1,2}$ are the diquark and antidiquark spins. The numerical analysis shows that this form factor can be approximated with high accuracy by the expression
\begin{equation}
 \label{eq:fr}
 F(r)=1-e^{-\xi r -\zeta r^2}.
\end{equation}
Such form factor smears the one-gluon exchange potential and removes
spurious singularities in the local relativistic quasipotential thus
allowing one to use it nonperturbatively to find the numerical
solution of the quasipotential equation. The masses and parameters of
light, heavy-light and doubly heavy diquarks are the same as in the
heavy baryons \cite{hbar,efgtetr,efgm,efgl} and are given in 
Tables~\ref{tab:dmass},\ref{tab:dqm}. As in the case of heavy baryons
we consider diquarks in the ground states only. In Fig.~\ref{fig:ud} we plot, as
an example, the form factors  $F(r)$ for the light scalar
$[u,d]$ and axial vector $\{u,d\}$ diquarks. For other diquarks the
form factors  $F(r)$ have similar form. 
As we see the functions $F(r)$ vanish in the limit $r\to 0$ and become
unity for large values of $r$. Such a behaviour can be easily understood
intuitively. At large distances a diquark can be well approximated by a
point-like object and its internal structure cannot be resolved. When
the distance to the diquark decreases the internal structure plays a more
important role. As the distance approaches zero, the interaction weakens and
turns to zero for $r=0$ since this point coincides with the center of
gravity of the two quarks forming the diquark. Thus the function $F(r)$
gives an important contribution to the short-range part of the
interaction of the light and  heavy diquark in the tetraquark and
can be neglected for the long-range (confining) interaction.

\begin{table}
  \caption{Masses $M$ and form factor  parameters of light and heavy-light
    diquarks. $S$ and $A$ 
    denote scalar and axial vector diquarks which are antisymmetric $[\cdots]$ and
    symmetric $\{\cdots\}$ in flavour, respectively. }
  \label{tab:dmass}
   \begin{ruledtabular}\begin{tabular}{ccccc}
Quark& Diquark&   $M$ &$\xi$ & $\zeta$
 \\
content &type & (MeV)& (GeV)& (GeV$^2$)  \\
\hline
$[u,d]$&S & 710 & 1.09 & 0.185  \\
$\{u,d\}$&A & 909 &1.185 & 0.365  \\
$[u,s]$& S & 948 & 1.23 & 0.225 \\
$\{u,s\}$& A & 1069 & 1.15 & 0.325\\
$\{s,s\}$ & A& 1203 & 1.13 & 0.280\\
$[c,u]$& $S$ & 1973& 2.55 &0.63  \\
$\{c,u\}$& $A$ & 2036& 2.51 &0.45  \\
$[c,s]$ & $S$& 2091& 2.15 & 1.05  \\
$\{c,s\}$& $A$ & 2158&2.12& 0.99 \\
$[b,u]$& $S$ & 5359 &6.10 & 0.55 \\
$\{b,u\}$& $A$ & 5381& 6.05 &0.35 \\
$[b,s]$ & $S$& 5462 & 5.70 &0.35 \\
    $\{b,s\}$& $A$ & 5482 & 5.65 &0.27\\
  \end{tabular}\end{ruledtabular}
\end{table}

\begin{table}
 \caption{Masses $M$ and form factor parameters of doubly heavy $QQ'$ diquarks. $S$ and $A$
 denote scalar and axial-vector diquarks, antisymmetric $[Q,Q']$ and
 symmetric $\{Q,Q'\}$ in flavour, respectively. }
\label{tab:dqm}
\begin{ruledtabular}
 \begin{tabular}{@{}cccccccc@{}}
Quark& Diquark&
\multicolumn{3}{c}{$Q=c$}
&\multicolumn{3}{c}{$Q=b$}
   \\ \cline{3-5} \cline{6-8}
   content &type & $M$ (MeV)&$\xi$ (GeV)&$\zeta$
(GeV$^2$) & $M$ (MeV)&$\xi$ (GeV)&$\zeta$ (GeV$^2$) \\ \hline
 $[Q,c]$ & $S$& & & & 6519 & 1.50
&0.59\\ $\{Q,c\}$& $A$& 3226& 1.30& 0.42 & 6526 & 1.50 &0.59\\
   $\{Q,b\}$& $A$& 6526 & 1.50 &0.59& 9778 & 1.30 &1.60\\
                                                   
 \end{tabular}
 \end{ruledtabular}
\end{table}

To calculate masses of the ground state and excited tetarquarks, we
substitute the diquark-antidiquark quasipotential (\ref{eq:pot}) in
the quasipotential equation (\ref{quas}) and solve the resulting
differential equation numerically in configuration space. It is
important to emphasize that all relativistic contributions to the
quasipotential are treated nonperturbatively. In the following
sections we present results of such calculations for the tetraquarks
containing two or four heavy quarks.  

\section{Heavy tetraquarks  $(Qq)(\bar Q\bar q')$ with hidden charm
  and bottom}
\label{sec:htQq}

First, we consider heavy tetraquarks with hidden charm and
bottom $(Qq)(\bar Q\bar q')$ ($Q=c$ or $b$, $q,q'=u,d,s$). They can
provide candidates for exotic charmonium-like (see
Table~\ref{ccccexp}) and bottomonium-like ( $Z_b(10610)$ and
$Z_b(10650)$) states observed experimentally.  

In the diquark-antidiquark picture of heavy tetraquarks
both scalar $S$ (antisymmetric in flavour
$(Qq)_{S=0}=[Qq]$) and axial vector $A$ (symmetric in flavour
$(Qq)_{S=1}=\{Qq\}$) diquarks are considered. Therefore we get the
following structure of the $(Qq)(\bar Q\bar q')$  ground ($1S$) states
($C$ is defined only for $q=q'$): 
\begin{itemize}
\item Two states with $J^{PC}=0^{++}$:
\begin{eqnarray*}
&&X(0^{++})=(Qq)_{S=0}(\bar Q\bar q')_{S=0}\\
&&X(0^{++}{}')=(Qq)_{S=1}(\bar Q\bar q')_{S=1}
\end{eqnarray*}
\item Three states with $J=1$:
\begin{eqnarray*}
&&X(1^{++})=\frac1{\sqrt{2}}[(Qq)_{S=1}(\bar Q\bar q')_{S=0}+(Qq)_{S=0}(\bar Q\bar
  q')_{S=1}]\\
&&X(1^{+-})=\frac1{\sqrt{2}}[(Qq)_{S=0}(\bar Q\bar q')_{S=1}-(Qq)_{S=1}(\bar Q\bar
  q')_{S=0}]\\
&&X(1^{+-}{}')=(Qq)_{S=1}(\bar Q\bar q')_{S=1}
\end{eqnarray*}
\item  One state with $J^{PC}=2^{++}$:
$$X(2^{++})=(Qq)_{S=1}(\bar Q\bar q')_{S=1}.$$
\end{itemize}
The orbitally excited ($1P,1D\dots$) states are
constructed analogously. As we find, a very rich spectrum of tetraquarks
emerges. However, the number of states in the considered
diquark-antidiquark picture is significantly less than in the genuine
four-quark approach.

\begingroup
\squeezetable
\begin{table}
  \caption{Masses of hidden charm diquark-antidiquark states
     (in MeV).  $S$ and $A$
    denote scalar and axial vector diquarks; $\cal S$ is the total
    spin of the diquark and antidiquark. ($C$ is defined only for $q=q'$).}
  \label{tab:ecmass}
\begin{ruledtabular}
\begin{tabular}{cccccc}
State& Diquark & &
\multicolumn{3}{c}{Tetraquark mass} 
  \\
  \cline{4-6}
$J^{PC}$ & content&$\cal S$& $cq\bar c\bar q$ &$cs\bar c\bar s$ & $cq\bar c\bar s$ \\
  \hline
  $1S$\\
$0^{++}$ & $S\bar S$&0 & 3812 & 4051 & 3922\\
$1^{+\pm}$& $(S\bar A\pm \bar S A)/\sqrt2$&1 & 3871& 4113 & 3982\\
$0^{++}$& $A\bar A$&0 & 3852 & 4110& 3967\\
$1^{+-}$& $A\bar A$&1 & 3890 & 4143& 4004\\
  $2^{++}$& $A\bar A$&2 & 3968 & 4209&4080\\
  \hline
$1P$\\
$1^{--}$ & $S\bar S$& 0 & 4244& 4466& 4350\\
$0^{-\pm}$ & $(S\bar A\pm \bar S A)/\sqrt2$&1&4269 & 4499&4381 \\
$1^{-\pm}$ & $(S\bar A\pm \bar S A)/\sqrt2$&1&4284 &4514& 4396 \\
$2^{-\pm}$ & $(S\bar A\pm \bar S A)/\sqrt2$&1&4315 &4543 &4426 \\
$1^{--}$& $A\bar A$ &0&4350& 4582& 4461\\
$0^{-+}$& $A\bar A$ & 1&4304 &4540& 4419\\
$1^{-+}$& $A\bar A$ & 1& 4345 &4578& 4458\\
$2^{-+}$& $A\bar A$ & 1& 4367 & 4598& 4478\\
$1^{--}$& $A\bar A$ & 2&4277 &4515& 4393\\
$2^{--}$& $A\bar A$ & 2& 4379& 4610& 4490\\
  $3^{--}$& $A\bar A$ & 2& 4381& 4612& 4492\\
  \hline
 $2S$\\
$0^{++}$ & $S\bar S$ & 0 & 4375 &4604&4481\\
$1^{+\pm}$ & $(S\bar A\pm \bar S A)/\sqrt2$& 1& 4431 & 4665& 4542\\
$0^{++}$& $A\bar A$ & 0 & 4434 &4680& 4547\\
$1^{+-}$& $A\bar A$ & 1 & 4461 &4703&4572\\
  $2^{++}$& $A\bar A$ & 2 & 4515& 4748& 4625\\
  \hline
  $1D$\\
$2^{++}$ & $S\bar S$& 0 & 4506& 4728& 4611\\
$1^{+\pm}$ & $(S\bar A\pm \bar S A)/\sqrt2$&1&4553 &4779& 4663\\
$2^{+\pm}$ & $(S\bar A\pm \bar S A)/\sqrt2$&1&4559 &4785& 4670\\
$3^{+\pm}$ & $(S\bar A\pm \bar S A)/\sqrt2$&1&4570 &4794& 4680\\
$2^{++}$& $A\bar A$ &0&4617& 4847&4727\\
$1^{+-}$& $A\bar A$ & 1&4604 &4835&4714\\
$2^{+-}$& $A\bar A$ & 1& 4616&4846&4726\\
$3^{+-}$& $A\bar A$ & 1& 4624&4852& 4733\\
$0^{++}$& $A\bar A$ & 2&4582 &4814& 4692\\
$1^{++}$& $A\bar A$ & 2&4593& 4825& 4703\\
$2^{++}$& $A\bar A$ & 2& 4610 & 4841& 4720\\
$3^{++}$& $A\bar A$ & 2& 4627 &4855& 4736\\
  $4^{++}$& $A\bar A$ & 2&4628& 4856& 4738\\
  \hline
$2P$\\
$1^{--}$ & $S\bar S$& 0 & 4666& 4884& 4767\\
$0^{-\pm}$ & $(S\bar A\pm \bar S A)/\sqrt2$&1&4684 & 4909&4792\\
$1^{-\pm}$ & $(S\bar A\pm \bar S A)/\sqrt2$&1&4702 & 4926& 4810\\
$2^{-\pm}$ & $(S\bar A\pm \bar S A)/\sqrt2$&1&4738 & 4960& 4845\\
$1^{--}$& $A\bar A$ &0&4765& 4991&4872\\
$0^{-+}$& $A\bar A$ & 1&4715& 4946& 4826\\
$1^{-+}$& $A\bar A$ & 1& 4760 &4987& 4867\\
$2^{-+}$& $A\bar A$ & 1& 4786& 5011& 4892\\
$1^{--}$& $A\bar A$ & 2&4687&4920&4799\\
$2^{--}$& $A\bar A$ & 2& 4797 &5022&4903\\
    $3^{--}$& $A\bar A$ & 2& 4804 &5030&4910\\
 \end{tabular}
\end{ruledtabular}
\end{table}
\endgroup

\begingroup
\squeezetable
\begin{table}
 \caption{Masses of hidden bottom tetraquark states (in MeV).}
  \label{tab:bmass}
\begin{ruledtabular}
  \begin{tabular}{cccccc}
State& Diquark & &
\multicolumn{3}{c}{Tetraquark mass} 
 \\
  \cline{4-6}
$J^{PC}$ & content&$\cal S$& $bq\bar b\bar q$ & $bs\bar b\bar s$ & $bq\bar b\bar s$ \\ 
\hline
$1S$\\
$0^{++}$& $S\bar S$&0  & 10471 & 10662 & 10572\\
$1^{+\pm}$& $(S\bar A\pm \bar S A)/\sqrt2$& 1& 10492& 10682 & 10593\\
$0^{++}$& $A\bar A$ &0& 10473 & 10671& 10584\\
$1^{+-}$& $A\bar A$ &1& 10494 & 10686& 10599\\
    $2^{++}$& $A\bar A$ &2& 10534 & 10716& 10628\\
    \hline
  $1P$\\
$1^{--}$ & $S\bar S$& 0 & 10807& 11002& 10907\\
$0^{-\pm}$ & $(S\bar A\pm \bar S A)/\sqrt2$&1&10820 & 11011&10917 \\
$1^{-\pm}$ & $(S\bar A\pm \bar S A)/\sqrt2$&1&10824 &11016& 10922 \\
$2^{-\pm}$ & $(S\bar A\pm \bar S A)/\sqrt2$&1&10834 &11026 &10932 \\
$1^{--}$& $A\bar A$ &0&10850& 11039& 10947\\
$0^{-+}$& $A\bar A$ & 1&10836 &11026& 10934\\
$1^{-+}$& $A\bar A$ & 1& 10847 &11037& 10945\\
$2^{-+}$& $A\bar A$ & 1& 10854 & 11044& 10952\\
$1^{--}$& $A\bar A$ & 2&10827 &11017& 10925\\
$2^{--}$& $A\bar A$ & 2& 10856& 11046& 10953\\
    $3^{--}$& $A\bar A$ & 2& 10858& 11048& 10956\\
    \hline
   $2S$\\
$0^{++}$ & $S\bar S$ & 0 & 10917 &11111&11018\\
$1^{+\pm}$ & $(S\bar A\pm \bar S A)/\sqrt2$& 1& 10939 & 11130& 11037\\
$0^{++}$& $A\bar A$ & 0 & 10942 &11133& 11041\\
$1^{+-}$& $A\bar A$ & 1 & 10951 &11142&11050\\
    $2^{++}$& $A\bar A$ & 2 & 10969& 11159& 11067\\
    \hline
    $1D$\\
$2^{++}$ & $S\bar S$& 0 & 11021& 11216& 11121\\
$1^{+\pm}$ & $(S\bar A\pm \bar S A)/\sqrt2$&1&11040 &11232& 11137\\
$2^{+\pm}$ & $(S\bar A\pm \bar S A)/\sqrt2$&1&11042 &11235& 11139\\
$3^{+\pm}$ & $(S\bar A\pm \bar S A)/\sqrt2$&1&11045 &11238& 11142\\
$2^{++}$& $A\bar A$ &0&11064& 11255&11162\\
$1^{+-}$& $A\bar A$ & 1&11060 &11251&11158\\
$2^{+-}$& $A\bar A$ & 1& 11064&11254&11161\\
$3^{+-}$& $A\bar A$ & 1& 11066&11257& 11164\\
$0^{++}$& $A\bar A$ & 2&11054 &11245& 11152\\
$1^{++}$& $A\bar A$ & 2&11057& 11248& 11155\\
$2^{++}$& $A\bar A$ & 2& 11062 & 11252& 11159\\
$3^{++}$& $A\bar A$ & 2& 11066 &11257& 11164\\
    $4^{++}$& $A\bar A$ & 2&11067& 11259& 11165\\
    \hline
$2P$\\
$1^{--}$ & $S\bar S$& 0 & 11122& 11316& 11221\\
$0^{-\pm}$ & $(S\bar A\pm \bar S A)/\sqrt2$&1&11134 & 11326&11232\\
$1^{-\pm}$ & $(S\bar A\pm \bar S A)/\sqrt2$&1&11139 & 11330&11236\\
$2^{-\pm}$ & $(S\bar A\pm \bar S A)/\sqrt2$&1&11148 & 11340&11245\\
$1^{--}$& $A\bar A$ &0&11163& 11353&11260\\
$0^{-+}$& $A\bar A$ & 1&11151& 11342& 11248\\
$1^{-+}$& $A\bar A$ & 1&11161 &11351& 11259\\
$2^{-+}$& $A\bar A$ & 1&11168& 11358& 11265\\
$1^{--}$& $A\bar A$ & 2&11143& 11333& 11241\\
$2^{--}$& $A\bar A$ & 2&11169 &11359& 11266\\
$3^{--}$& $A\bar A$ & 2&11172 &11362& 11269\\
\end{tabular}
\end{ruledtabular}
\end{table}
\endgroup

The diquark-antidiquark model of heavy tetraquarks predicts
 the existence of the flavour 
$SU(3)$ nonet of  states with hidden
charm or beauty ($Q=c,b$): four tetraquarks
[$(Qq)(\bar Q\bar q)$, $q=u,d$] with neither open 
or hidden strangeness, which have
electric charges 0 or $\pm 1$ and isospin 0 or 1; 
four tetraquarks [$(Qs)(\bar Q\bar q)$
and  $(Qq)(\bar Q\bar s)$, $q=u,d$] with open strangeness ($S=\pm 1$),
which have electric charges 0 or $\pm 1$ and isospin $\frac12$; 
one tetraquark
$(Qs)(\bar Q\bar s)$ with hidden strangeness and zero electric
charge. 
Since we neglect  in our model the mass difference of $u$ and
$d$ quarks and electromagnetic interactions, the corresponding tetraquarks
will be degenerate in mass. A more 
detailed analysis \cite{mppr}
predicts that the tetraquark mass differences can be of a few MeV so
that the
isospin invariance is broken for the $(Qq)(\bar Q\bar q)$ mass
eigenstates and thus in their strong decays.

Masses of the ground,
orbitally  and radially excited states of   heavy tetraquarks  were
calculated in Refs.~\cite{efgtetr,efgextetr,efgb} and we give them in
Tables~\ref{tab:ecmass} and  \ref{tab:bmass}. Note that most of the ground
tetraquark states are predicted to lie 
either above or only slightly below corresponding open charm and bottom
thresholds. For the excited states we consider excitations only of the diquark-antidiquark system. A very rich spectrum of excited tetraquark
states is obtained.

\begin{table}
  \caption{Masses of hidden
   charm diquark-antidiquark states   (in MeV) and
   possible experimental candidates.}\label{qqcomp}
 \begin{ruledtabular}
\begin{tabular}{@{}c@{ } c@{ }c @{ }c@{ } c@{ } c@{ }c@{ } c @{ }c@{}}
State&Diquark&
\multicolumn{3}{c}{Theory}
& \multicolumn{2}{c}{Experiment} &Theory
  \\ \cline{3-5} \cline{6-7}
$J^{PC}$&content &$cq\bar c\bar q$ & $cs\bar c\bar s$ &$cq\bar c\bar s$
&state& mass &$bq\bar b\bar q$\\
\hline
$1S$\\ 
$1^{++}$&$(S\bar A+ \bar S A)/\sqrt2$ & 3871&& &$X$(3872)
&$3871.69\pm0.17$ 
                 &10492\\
  $1^{+-}$&$A\bar A$& 3890&& &$Z_c$(3900)& $3888.4\pm2.5$& 10494\\
$1^{+}$&$(S\bar A- \bar S A)/\sqrt2$ &&&3982&$Z_{cs}(3985)$&$3982.5^{+1.8}_{-2.6}\pm2.1$&10593\\  
$1^{++}$&$(S\bar A+
  \bar S A)/\sqrt2$ & &   4113& &$X$(4140)&$4146.8\pm2.4$& 10682\\
$2^{++}$&$A\bar A$& 3968 &&
&$?^{??}$ $X$(3940)&$3942^{+7}_{-6}\pm6$&10534\\
$1P$\\
$1^{--}$&$S\bar S$&4244 & &&$Y$(4230)
&$4218.7\pm2.8$& 10807\\
$1^{--}$&$A\bar A$&4277 & &&$Y$(4260)
&$ {4230}\pm 8 $& 10827\\                
$0^{--}$&$
  (S\bar A- \bar S A)/\sqrt2$&4269&&&$Z_c$(4240)&$4239\pm18^{+45}_{-10}$
 & 10820\\
$\begin{array}{l}0^{-+}\\1^{-+} \end{array}$&$\begin{array}{c}(S\bar
  A+ \bar S A)/\sqrt2\\ (S\bar
  A+ \bar S
                                                A)/\sqrt2 \end{array}$&$\left.\begin{array}{l}4269\\4284 \end{array}\right\}$& &&$?^{?+}$ $Z_c$(4250) &
4248$^{+44+180}_{-29-35}$&$\begin{array}{l}10820\\10824 \end{array}$\\
$1^{--}$&$A\bar A$& 4350&&   &$Y$(4360) &
$4368\pm13 $& 10850\\
$2S$\\
$1^{+-}$&$\begin{array}{c}(S\bar A - \bar S A)/\sqrt2\\A\bar A \end{array}$&$\left.\begin{array}{l}4431\\4461\end{array}\right\}$&& 
&$Z_c$(4430)&$4478^{+15}_{-18}$ & $\begin{array}{l}10939\\10951 \end{array}$\\
$0^{++}$&$S\bar S$&&4604&& $X$(4500)&$4506\pm11^{+12}_{-15}$&11111 \\
                  $0^{++}$&$A\bar A$&&4680&&$X$(4700)  &$4704\pm10^{+14}_{-24}$ &11133\\
 $2^{++}$&  $A\bar A$&&4748&&$?^{?+}$$X(4740)$&$4741\pm6\pm6$&11159\\               
$2P$\\
$1^{--}$&$S\bar S$&4666&  && $Y$(4660)&
$4633\pm7$&11122\\ 
\end{tabular}
\end{ruledtabular}
\end{table}

In Table~\ref{qqcomp} we compare the predicted masses of tetraquarks
with hidden charm with available experimental data, listed in Table~\ref{ccccexp}, and give possible
tetraquarks candidates. For the exotic charmonium-like states we get
the following results.
The predicted mass of the ground state $1^{++}$ neutral charm
tetraquark state coincides with the measured mass of $X(3872)$ . Then the charged
$Z_c(3900)$  can be its $1^{+-}$ partner state, composed from axial
vector diquark ($A$) and axial vector antidiquark ($\bar A$), and $Z_c(4430)$ is its first radial
excitation. Indeed, the predicted masses of these states are within
experimental error bars. From its value of the mass,  $X(3940)$ with unmeasured quantum numbers
could be $2^{++}$ of the $A\bar A$ tetraquark.
The charged $Z_c(4020)$, $Z_c(4050)$, $Z_c(4055)$, $Z_c(4100)$
and $Z_c(4200)$ have masses inconsistent with our results. They could
be, e.g., the hadro-charmonium or molecular states.
The charged $Z_c(4240)$ can be the
$0^{--}$ state of $1P$-wave tetraquark, composed from scalar ($S$) and axial
vector ($A$) diquark-antidiquark combinations, while controversial
$Z_c(4250)$ with the unmeasured parity  and poorly determined
mass could be its  $0^{-+}$ or $1^{-+}$ partner.
The vector $Y(4230)$, $Y(4260)$ and $Y(4360)$ 
can be the $1^{--}$ $1P$-wave  
tetraquark states, composed from $S\bar S$ and $A\bar A$ diquarks,
respectively,  while $Y(4660)$ corresponds to the $2P$-wave state of the  $S\bar S$
tetraquark. We have no tetraquark candidate for the $Y(4390)$ state.

Now we discuss the exotic charmonium-like states observed in the
$J/\psi\phi$ mass spectrum.  The axial vector $X(4140)$ can be the $[cs][\bar
c\bar s]$ ground state tetraquark with $1^{++}$, composed form a scalar ($S$) and axial
vector ($A$) diquark-antidiquark combinations, while the scalar $X(4500)$
and $X(4700)$ can correspond to the first radially excited  $0^{++}$
tetraquarks, composed from the $S\bar S$ and $A\bar A$, respectively.
If $X(4740)$, very recently observed   by LHCb \cite{Aaij:2020tzn}, is different from
$X(4700)$ it can be the $2S$ excitation of the  $A\bar A$ tetraquark with
$2^{++}$. We do not have the tetraquark candidate for the $X(4274)$.
The mass of the very recently observed \cite{Zcs} charged state with
open strangeness $Z_{cs}(3985)^-$ coincides with our prediction for
the $1^+$ state composed from scalar and axial vector diquarks $(S\bar A- \bar S A)/\sqrt2$ .   
It is important to point out that most of the exotic charmonium-like
states were discovered experimentally after our predictions.

In the exotic botomonium-like sector 
we do not have tetraquark
candidates for the charged $Z_b(10610)$ and $Z_b(10650)$, which are
probably a molecular states.
 The ground states of the
tetraquarks with hidden bottom are predicted to have masses below the 
open bottom threshold and thus they should be narrow states.
 In the last column of Table~\ref{qqcomp} the predictions for the masses of bottom counterparts to
the hidden charm tetraquark candidates are given.

\section{ Doubly heavy tetraquarks with open charm and bottom $(QQ')(\bar q\bar
  q')$.}
\label{sec:dht}

The doubly heavy $(QQ')(\bar q\bar
q')$ tetraquark ($Q,Q'=b,c$ and $q,q'=u,d,s$) is considered as the bound
system of the heavy diquark ($QQ'$) and light antidiquark ($\bar
q\bar q'$).
It is important to investigate the possible stability of the
$(QQ')(\bar q\bar q')$ tetraquarks since they are explicitly
exotic states with the heavy flavour number equal to 2. Thus,
their observation would be a direct proof of the existence of the
multiquark states. Estimates of the production rates of such
tetraquarks indicate that they could be produced and detected at
present and future facilities.

\begin{table}
  \caption{Masses $M$ of heavy-diquark ($QQ'$)--light-antidiquark ($\bar q\bar q$)
    states. $T$ is the lowest threshold for decays into two
    heavy-light ($Q\bar q$) mesons and $\Delta=M-T$. All values are
    given in MeV.}
  \label{tab:QQqqmass}
  \begin{ruledtabular}
\begin{tabular}{@{}ccccccccccc@{}}
System&State&
\multicolumn{3}{c}{$Q=Q'=c$}&
\multicolumn{3}{c}{$Q=Q'=b$}&
\multicolumn{3}{c}{$Q=c$, $Q'=b$}
  \\
  \cline{3-5} \cline{6-8} \cline{9-11}
&$I(J^{P})$ & $M$ &$T$ & $\Delta$ & $M$ &$T$ & $\Delta$ & $M$ &$T$ & $\Delta$\\
\hline
$(QQ')(\bar u\bar d)$&&&&&&&&&&\\
&$0(0^+)$ &        &      &    &       &       &        & 7239 & 7144
& 95\\
&$0(1^{+})$ & 3935 & 3871 & 64 & 10502 & 10604 & $-102$ & 7246 & 7190
& 56\\
&$1(1^+)$ &        &      &    &       &       &        & 7403 & 7190
& 213\\
&$1(0^{+})$ & 4056 & 3729 & 327& 10648 & 10558 & 90 & 7383 & 7144 & 239\\
&$1(1^{+})$ & 4079 & 3871 & 208& 10657 & 10604 & 53 & 7396 & 7190 & 206\\
&$1(2^{+})$ & 4118 & 4014 & 104& 10673 & 10650 & 23 & 7422 & 7332 & 90\\
$(QQ')(\bar u\bar s)$&&&&&&&&&&\\
&$\frac12(0^+)$ &        &      &    &       &       &        & 7444 & 7232
& 212\\
&$\frac12(1^{+})$ & 4143 & 3975 & 168 & 10706 & 10693 & 13 & 7451 &
7277 & 174 \\
&$\frac12(1^{+})$ &   &   &   &   &  &  & 7555 & 7277 & 278\\
&$\frac12(0^{+})$ & 4221 & 3833 & 388 & 10802 & 10649 & 153 & 7540 &
7232 & 308 \\
&$\frac12(1^{+})$ & 4239 & 3975 & 264 & 10809 & 10693 & 116 & 7552 &
7277 & 275 \\
&$\frac12(2^{+})$ & 4271 & 4119 & 152 & 10823 & 10742 & 81 & 7572 & 7420& 152\\
$(QQ')(\bar s\bar s)$&&&&&&&&&&\\
&$0(1^{+})$ &      &      &    &       &       &        & 7684 & 7381
& 303\\
&$0(0^{+})$ & 4359 & 3936 & 423 & 10932 & 10739 & 193 & 7673 & 7336 & 337\\
&$0(1^{+})$ & 4375 & 4080 & 295 & 10939 & 10786 & 153 & 7683 & 7381 &
302\\
&$0(2^{+})$ & 4402 & 4224 & 178 & 10950 & 10833 & 117 & 7701 & 7525 &
176\\
\end{tabular}
\end{ruledtabular}
\end{table}

We calculated the masses $M$ of the ground states ($1S$) of doubly heavy
tetraquarks with open charm and/or bottom composed from the heavy
diquark, containing two heavy quarks ($QQ'$, $Q,Q'=b,c$), and the
light antidiquark ($\bar q\bar q'$, $q,q'=u,d,s$) in Ref.~\cite{efgl}. They are presented in
Table~\ref{tab:QQqqmass}. In this table we give the values of the
lowest thresholds $T$ for decays into two corresponding
heavy-light mesons [$(Q\bar
q)=D^{(*)},D_s^{(*)},B^{(*)},B_s^{(*)}$], which were calculated
using the measured masses of these mesons \cite{pdg}. We also show
values of the difference of the tetraquark and threshold masses
$\Delta=M-T$. If this quantity is negative, then the tetraquark
lies below the threshold of the decay into mesons with open
flavour and thus should be a narrow state which can be detected
experimentally. The states with small positive values of $\Delta$
could be also observed as resonances, since their decay rates will
be suppressed by the phase space. All other states are expected to
be very broad and thus unobservable. We find that the only
tetraquark which lies considerably below threshold is the $0(1^+)$
state of $(bb)(\bar u\bar d)$. All other $(QQ')(\bar q \bar q')$
tetraquarks are predicted to lie either close to ($1(2^+)$ and $1(1^+)$
states of $(bb)(\bar u\bar d)$, $\frac12(1^+)$
state of $(bb)(\bar u\bar s)$,  $0(1^+)$
state of $(cb)(\bar u\bar d)$, $0(1^+)$
state of $(cc)(\bar u\bar d)$) or significantly
above corresponding thresholds. Note that our predictions are in
accord with the recent  lattice QCD calculations \cite{latt1,latt2,latt3,latt4}, which find that only
$J^P=1^+, I=0$ doubly bottom tetraquarks have masses below the
corresponding two-meson thresholds. This conclusion is also supported
by the heavy quark symmetry \cite{eq} and quark model relations
\cite{kr}.

It is evident from the results
presented in Table~\ref{tab:QQqqmass} that the heavy tetraquarks
have increasing chances to be below the open flavour threshold and,
thus have a narrow width, with the increase of the ratio of the
heavy diquark mass to the light antidiquark mass. 

It is important to note that the comparison of the masses of doubly heavy
tetraquarks given in Table~\ref{tab:QQqqmass} with our  predictions for
the masses of hidden charm and bottom tetraquarks in
Tables~\ref{tab:ecmass}, \ref{tab:bmass} \cite{efgtetr,efgextetr,efgb}
shows that the $(QQ')(\bar q\bar q')$ states are, in general,
heavier than the corresponding $(Qq)(\bar Q'\bar q')$ ones. This
result has the following explanation. Although the relation
 $M_{QQ}+M^S_{qq}\le 2M_{Qq}$ holds between diquark
masses, the binding energy in the heavy-light diquark ($Qq$)--
heavy-light antidiquark ($\bar Q\bar q$) bound system is
significantly larger than in the corresponding heavy diquark
($QQ$)--light antidiquark ($\bar q\bar q$) one. This fact is well
known from the meson spectroscopy, where heavy quarkonia $Q\bar Q$
are more tightly bound than heavy-light mesons $Q\bar q$. For
instance, we found that some of the $(cu)(\bar c\bar u)$
tetraquarks lie below open charm thresholds while all ground-state
$(cc)(\bar u \bar d)$ tetraquarks are found to be above such
thresholds.

\section{Heavy tetraquarks  $(cq)(\bar b\bar q')$ with open charm and
  bottom}
\label{sec:htof}

\begin{table}
 \caption{Masses $M$ of diquark ($cq'$)--antidiquark ($\bar b\bar q$)
 states. $T$ is the lowest threshold for decays into two
 heavy-light ($Q\bar q$) mesons and $\Delta=M-T$; $T'$ is the
 threshold for decays into the $B_c^{(*)}$ and a light meson ($q'\bar q$), and
 $\Delta'=M-T'$. All values are
 given in MeV.}
\label{tab:cqbqmass}
\begin{ruledtabular}
 \begin{tabular}{cccccccccccc@{}}
System&State&
\multicolumn{5}{c}{$q'=u$}&
\multicolumn{5}{c}{$q'=s$}
\\ \cline{3-7} \cline{8-12}
&$J^{P}$ & $M$ &$T$ & $\Delta$&$T'$ & $\Delta'$ & $M$ &$T$ &
                                                             $\Delta$&$T'$ & $\Delta'$ \\
   \hline
   $(cq')(\bar b\bar u)$&&&&&&&&&&\\ &$0^+$
& 7177 & 7144 & 33& 6818&359 & 7294 & 7232 & 62& 6768 & 526 \\
&$1^{+}$ & 7198 & 7190 & 8 & 6880 & 318 & 7317 & 7277 & 40 & 6820
                                                      & 497\\
&$1^+$ & 7242 & 7190 & 52 & 6880 & 362 & 7362 & 7277 & 85
                                                & 6820 & 542\\
&$0^{+}$ & 7221 & 7144 & 77 & 6818& 403 & 7343 &
                                                 7232 & 111 & 6768 & 575\\
&$1^{+}$ & 7242 & 7190 & 52 & 6880 & 362
                    & 7364 & 7277 & 87 & 6820 & 544\\
&$2^{+}$ & 7288 & 7332 & $-44$
         &7125 &163 & 7406 & 7420 & $-14$ & 7228 & 178\\
   $(cq')(\bar b\bar s)$&&&&&&&&&&\\
&$0^+$ & 7282 & 7247 & 35 & 6768 &514 & 7398 & 7336 & 62&
                                                          6818& 580 \\
&$1^{+}$ & 7302 & 7293 & 9 & 6820& 482 & 7418 & 7381
                                    & 37 &6880 & 538\\
&$1^+$ & 7346 & 7293 & 53 & 6820& 526 & 7465 &
                                               7381 & 84 & 6880 & 585\\
&$0^{+}$ & 7325 & 7247 & 78 & 6768 & 557
                    & 7445 & 7336 & 109& 6818 & 627\\
&$1^{+}$ & 7345 & 7293 & 52&
                             6820& 525& 7465 & 7381 & 84& 6880& 585 \\
&$2^{+}$ & 7389 & 7437 &
$-48$& 7228& 161& 7506 & 7525 &$-19$& 7352 & 154\\
 \end{tabular}
 \end{ruledtabular}
\end{table}

The $(cq)(\bar b\bar q')$ tetraquark is considered to be the
bound state of the heavy-light diquark ($cq$) and antidiquark
($\bar b\bar q')$. 
In Table~\ref{tab:cqbqmass} the calculated masses $M$ of the ground
states of heavy
tetraquarks with open charm and bottom, composed from a $(cq)$ diquark and a $(\bar b\bar q)$
antidiquark, are presented \cite{efgl}. We also give the lowest
thresholds $T$ for decays into heavy-light mesons as well as
thresholds $T'$ for decays into the $B_c^{(*)}$ and light ($q'\bar
q$) mesons and $\Delta^{(')}=M-T^{(')}$.\footnote{For the
non-strange $(cq)(\bar b\bar q)$ tetraquarks we give thresholds
$T'$ for decays of the $I=0$ states into $B_c^{(*)}$ and $\eta$ or
$\omega$. These states should be more stable than the $I=1$ ones,
since their decays to $B_c^{(*)}$ and
 $\pi$ violate isospin.} We find that only $2^+$ states of $(cq')(\bar
b \bar q)$ have negative values of $\Delta$ and thus they should
be stable with respect to decays into heavy-light ($B$ and $D$)
mesons. The predicted masses of lowest $1^+$ states of $(cu)(\bar
b \bar u)$ and $(cu)(\bar b \bar s)$ tetraquarks lie only slightly
above the corresponding thresholds $T$. However, all $(cq)(\bar
b\bar q)$ tetraquarks are found to be significantly above the
thresholds $T'$ for decays into the $B_c^{(*)}$ and light ($q'\bar
q$) mesons. Nevertheless, the wave function of the spatially
extended $(cq)(\bar b\bar q)$ tetraquark would have little overlap
with the wave function of the compact $B_c$ meson, thus substantially suppressing the decay rate
in this channel. Therefore the above-mentioned $(cq)(\bar b\bar
q)$ tetraquark states which are below the $BD$ threshold have good
chances to be rather narrow and could be detected experimentally.

\section{$QQ\bar Q\bar Q$ tetraquarks}
\label{sec:fht}

The exotic $QQ\bar Q\bar Q$ states  consisting of heavy
quarks ($Q=c$ and/or $b$) only are of special interest, since their nature
can be determined more easily than in the case of exotic charmonium
and bottomonium-like states. They should be predominantly compact
tetraquarks. Indeed, a molecular configuration is 
unlikely. Only heavy $Q\bar Q$ mesons can be exchanged between
constituents in such a molecule, and the arising Yukawa-type potential
is not strong enough to provide binding. Soft gluons can be exchanged
between two heavy quarkonia, leading to the so-called QCD van der
Waals force. Such a force is known to be attractive, though whether it
is strong enough to form a bound state remains unclear. The hadroquarkonium picture
is not applicable. Thus, the diquark
($QQ$)-antidiquark ($\bar Q\bar Q$) configuration is preferable.

 \begin{table}
 \caption{Masses $M$ of the neutral heavy diquark ($QQ$)-antidiquark ($\bar Q\bar Q$)
 states. $T$ is the threshold for the decays into two
 heavy-($Q\bar Q$) mesons and $\Delta=M-T$. All values are
 given~in~MeV.}
 \label{tab:QQmass}\begin{ruledtabular}
 \begin{tabular}{  c c c c c c c }
Composition &$d\bar d$  & $J^{PC}$ & $M$ & Threshold& $T$ & $\Delta$
\\
\hline
 \multirow{4}{*}{$cc \bar c \bar c$} & \multirow{4}{*}{$A \bar A$} & \multirow{2}{*}{$0^{++}$} & \multirow{2}{*}{6190} & $\eta_{c}(1S)\eta_{c}(1S)$ & 5968 & 222
\\
\cline{5-7}
  & & & &  $J/\psi(1S)J/\psi(1S)$ & 6194 & -4
\\
\cline{3-7}
  & & $1^{+-}$ & 6271 & $\eta_{c}(1S)J/\psi(1S)$ & 6081 & 190
\\
\cline{3-7}
  & & $2^{++}$ & 6367 & $J/\psi(1S)J/\psi(1S)$ & 6194 & 173
\\

\cline{1-7}
  \multirow{4}{*}{$bb \bar b \bar b$} & \multirow{4}{*}{$A \bar A$} & \multirow{2}{*}{$0^{++}$} & \multirow{2}{*}{19314} & $\eta_{b}(1S)\eta_{b}(1S)$ & 18797 & 517
\\
\cline{5-7}
  & & & &  $\Upsilon(1S)\Upsilon(1S)$ & 18920 & 394
\\
\cline{3-7}
  & & $1^{+-}$ & 19320 & $\eta_{b}(1S)\Upsilon(1S)$ & 18859 & 461
\\
\cline{3-7}
  & & $2^{++}$ & 19330 & $\Upsilon(1S)\Upsilon(1S)$ & 18920 & 410
   \\
\end{tabular}
\end{ruledtabular}
\end{table}

The calculated masses $M$ of the ground states \cite{fgs} of the neutral $QQ'\bar Q\bar Q'$ tetraquarks  composed of the heavy diquark ($QQ'$, $Q,Q'=b,c$), and heavy antidiquark ($\bar Q\bar Q'$) are given in
Tables~\ref{tab:QQmass},\ref{tab:QQmass1}. The masses of the charged heavy $QQ'\bar Q\bar Q'$ tetraquarks are presented in Table~\ref{tab:QQQQ'mass}. In these tables we give the values of the
lowest thresholds $T$ for decays into two corresponding
heavy mesons [$(Q\bar Q)$][$(Q'\bar Q')$] or [$(Q\bar Q')$][$(Q'\bar Q)$], which were calculated
using the measured masses of these mesons \cite{pdg}. We also show
values of the difference of the tetraquark and threshold masses,
$\Delta=M-T$. If this quantity is negative, then the tetraquark
lies below the threshold of the fall-apart decay into two mesons  and thus should be a narrow state. The states with small positive values of $\Delta$
could be also observed as resonances, since their decay rates will
be suppressed by the phase space. All other states are expected to
be broad and thus  difficult to observe.

\begin{table}
 \caption{Masses $M$ of the neutral heavy diquark ($cb$)-antidiquark ($\bar c\bar b$)
 states. $T$ is the threshold for the decays into two
 heavy-($Q\bar Q'$) mesons and $\Delta=M-T$. All values are
 given~in~MeV.}
 \label{tab:QQmass1}\begin{ruledtabular}
  \begin{tabular}{  c c c c c c c }
Composition &$d\bar d$  & $J^{PC}$ & $M$ & Threshold& $T$ & $\Delta$
\\
\hline
\cline{1-7}

  \multirow{21}{*}{$cb \bar c \bar b$} & \multirow{10}{*}{$A \bar A$} & \multirow{4}{*}{$0^{++}$} & \multirow{4}{*}{12813} & $\eta_{c}(1S)\eta_{b}(1S)$ & 12383 & 430
\\
\cline{5-7}
  & & & & $J/\psi(1S)\Upsilon(1S)$ & 12557 & 256
\\
\cline{5-7}
  & & & & $B_{c}^{\pm}  B_{c}^{\mp}$ & 12550 & 263
\\
\cline{5-7}
  & & & & $B_{c}^{*\pm} B_{c}^{*\mp}$ & 12666 & 147
\\
\cline{3-7}
  & & \multirow{4}{*}{$1^{+-}$} & \multirow{4}{*}{12826} & $\eta_{c}(1S)\Upsilon(1S)$ & 12444 & 382
\\
\cline{5-7}
  & & & & $J/\psi(1S)\eta_{b}(1S)$ & 12496 & 330
\\
\cline{5-7}
  & & & & $B_{c}^{\pm}  B_{c}^{*\mp}$ & 12608 & 218
\\
\cline{5-7}
  & & & & $B_{c}^{* \pm}  B_{c}^{* \mp}$ & 12666 & 160
\\
\cline{3-7}
  & & \multirow{2}{*}{$2^{++}$} & \multirow{2}{*}{12849} & $J/\psi(1S)\Upsilon(1S)$ & 12557 & 292
\\
\cline{5-7}
  & & & & $B_{c}^{* \pm}  B_{c}^{*\mp}$ & 12666 & 183
\\
\cline{2-7}
  & \multirow{7}{*}{$\frac{1}{\sqrt{2}}(A \bar S \pm S \bar A)$} & \multirow{3}{*}{$1^{++}$} & \multirow{4}{*}{12831} & $J/\psi(1S)\Upsilon(1S)$ & 12557 & 274
\\
\cline{5-7}
  & & & & $B_{c}^{\pm} B_{c}^{*\mp}$ & 12608 & 223
\\
\cline{5-7}
  & & & & $B_{c}^{*\pm}  B_{c}^{*\mp}$ & 12666 & 165
\\
\cline{3-7}
  & & \multirow{4}{*}{$1^{+-}$} &  \multirow{4}{*}{12831} & $\eta_{c}(1S)\Upsilon(1S)$ & 12444 & 387
\\
\cline{5-7}
  & & & & $J/\psi(1S)\eta_{b}(1S)$ & 12496 & 335
\\
\cline{5-7}
  & & & & $B_{c}^{\pm} B_{c}^{*\mp}$ & 12608 & 223
\\
\cline{5-7}
  & & & & $B_{c}^{*\pm}  B_{c}^{*\mp}$ & 12666 & 165
\\
\cline{2-7}
  & \multirow{4}{*}{$S \bar S$} & \multirow{4}{*}{$0^{++}$} & \multirow{4}{*}{12824} & $\eta_{c}(1S)\eta_{b}(1S)$ & 12383 & 441
\\
\cline{5-7}
  & & & & $J/\psi(1S)\Upsilon(1S)$ & 12557 & 267
\\
\cline{5-7}
  & & & & $B_{c}^{\pm}  B_{c}^{\mp}$ & 12550 & 274
\\
\cline{5-7}
  & & & & $B_{c}^{*\pm}  B_{c}^{*\mp}$ & 12666 & 158
                  \\
\end{tabular}\end{ruledtabular}
\end{table}

\begin{table}
 \caption{Masses $M$ of the charged heavy diquark--antidiquark 
 states. $T$ is the threshold for the decays into two
 heavy ($Q\bar Q'$) mesons and $\Delta=M-T$. All values are
 given~in~MeV.}
 \label{tab:QQQQ'mass}
 \begin{ruledtabular}\begin{tabular}{  c c c c c c c }
Composition &$d\bar d$  & $J^{P}$ & $M$ & Threshold& $T$ & $\Delta$
\\
\hline
 \multirow{9}{*}{$cc \bar c \bar b, cb \bar c \bar c$} & \multirow{6}{*}{$A \bar A$} & \multirow{2}{*}{$0^{+}$} & \multirow{2}{*}{9572} & $\eta_{c}(1S)B_{c}^{\pm}$ & 9259 & 313
\\
\cline{5-7}
  & & & &  $J/\psi(1S)B_{c}^{* \pm}$ & 9430 & 142
\\
\cline{3-7}
  & & \multirow{3}{*}{$1^{+}$} & \multirow{3}{*}{9602} & $\eta_{c}(1S)B_{c}^{* \pm}$ & 9317 & 285
\\
\cline{5-7}
  & & & & $J/\psi(1S)B_{c}^{\pm}$ & 9372 & 230
\\
\cline{5-7}
  & & & & $J/\psi(1S)B_{c}^{* \pm}$ & 9430 & 172
\\
\cline{3-7}
  & & $2^{+}$ & 9647 & $J/\psi(1S)B_{c}^{* \pm}$ & 9430 & 217
\\
\cline{2-7}
  & \multirow{3}{*}{$A \bar S$,  $ S \bar A$} & \multirow{3}{*}{$1^{+}$} & \multirow{3}{*}{9619} & $\eta_{c}(1S)B_{c}^{* \pm}$ & 9317 & 302
\\
\cline{5-7}
  & & & & $J/\psi(1S)B_{c}^{\pm}$ & 9372 & 247
\\
\cline{5-7}
  & & & & $J/\psi(1S)B_{c}^{* \pm}$ & 9430 & 189
\\
\cline{1-7}
  \multirow{5}{*}{$cc \bar b \bar b, bb \bar c \bar c$} & \multirow{5}{*}{$A \bar A$} & \multirow{2}{*}{$0^{+}$} & \multirow{2}{*}{12846} & $B_{c}^{\pm}B_{c}^{\pm}$ & 12550 & 296
\\
\cline{5-7}
  & & & & $B_{c}^{* \pm}B_{c}^{* \pm}$ & 12666 & 180
\\
\cline{3-7}
  & & \multirow{2}{*}{$1^{+}$} & \multirow{2}{*}{12859} & $B_{c}^{\pm} B_{c}^{* \pm}$ & 12608 & 251
\\
\cline{5-7}
  & & & & $B_{c}^{* \pm}B_{c}^{* \pm}$ & 12666 & 193
\\
\cline{3-7}
  & & \multirow{1}{*}{$2^{+}$} & \multirow{1}{*}{12883} & $B_{c}^{* \pm}B_{c}^{* \pm}$ & 12666 & 217
\\
\cline{1-7}
  \multirow{9}{*}{$cb \bar b \bar b, bb \bar c \bar b$} & \multirow{6}{*}{$A \bar A$} & \multirow{2}{*}{$0^{+}$} & \multirow{2}{*}{16109} & $B_{c}^{\pm}\eta_{b}(1S)$ & 15674 & 435
\\
\cline{5-7}
  & & & &  $B_{c}^{* \pm}\Upsilon(1S)$ & 15793 & 316
\\
\cline{3-7}
  & & \multirow{3}{*}{$1^{+}$} & \multirow{3}{*}{16117} & $B_{c}^{\pm}\Upsilon(1S)$ & 15735 & 382
\\
\cline{5-7}
  & & & & $B_{c}^{* \pm}\eta_{b}(1S)$ & 15732 & 385
\\
\cline{5-7}
  & & & & $B_{c}^{* \pm}\Upsilon(1S)$ & 15793 & 324
\\
\cline{3-7}
  & & $2^{+}$ & 16132 & $B_{c}^{* \pm}\Upsilon(1S)$ & 15793 & 339
\\
\cline{2-7}
  & \multirow{3}{*}{$S \bar A$, $A \bar S  $} & \multirow{3}{*}{$1^{+}$} & \multirow{3}{*}{16117} & $B_{c}^{\pm}\Upsilon(1S)$ & 15735 & 382
\\
\cline{5-7}
 & & & & $B_{c}^{* \pm}\eta_{b}(1S)$ & 15732 & 385
\\
\cline{5-7}
  & & & & $B_{c}^{* \pm}\Upsilon(1S)$ & 15793 & 324
\\
\end{tabular}\end{ruledtabular}
\end{table}

From these tables we see that the predicted masses of almost all $QQ\bar
Q\bar Q$ tetraquarks lie significantly higher than the thresholds of
the fall-apart decays to the lowest allowed two quarkonium states. All
these states should be broad, since they can decay to corresponding
quarkonium states through quark and antiquark rearrangements,  and
these decays  are
not suppressed either dynamically or 
kinematically. This conclusion is in accord with the current experimental
data. Indeed, the LHCb \cite{Aaij:2018zrb} and CMS \cite{Sirunyan:2020txn} Collaborations have not
observed narrow beautiful tetraquarks in the $\Upsilon(1S)$-pair
production. Note that the lattice nonrelativistic QCD \cite{lattnrqcd}
calculations did not find a signal for the $bb\bar b\bar b$
tetraquarks below the lowest noninteracting two-bottomonium threshold.
On the other hand the broad structure near the di-$J/\psi$
mass threshold very recently observed by the LHCb \cite{Aaij:2020fnh}
can correspond to the $2^{++}$ state of the $cc\bar c\bar c$
tetraquark, with a mass predicted to be 6367 MeV. The narrow
structure, $X(6900)$ \cite{Aaij:2020fnh}, could be the orbital or radial excitation of
this tetraquark.  Such excited states can be
narrow despite the large phase space since it will be necessary in the
fall-apart process to overcome the suppression either due to the
centrifugal barrier for the orbital excitations or due to the presence
of the nodes in the wave function of the radially excited state.    To
test this possibility we calculated masses of excited $cc\bar c\bar c$ tetraquarks. Both radial and orbital excitations only between the axial vector $\{c,c\}$
diquark and the axial vector $\{\bar c,\bar c\}$ antidiquark were considered. 

\begin{table}
  \caption{Masses $M$ of  $cc\bar c\bar c$ tetraquarks (in
    MeV); ${\cal S}$ is the total spin of the diquark and antidiquark.}\label{ccccexc}
   \begin{ruledtabular}
\begin{tabular}{cccccccccccc}
  State&$J^{PC}$& ${\cal S}$ & $M$&State&$J^{PC}$& ${\cal S}$ & $M$&State&$J^{PC}$& ${\cal S}$ & $M$\\
  \hline
  1S&&&& 1P&&&&1D\\
       &$0^{++}$&0&6190& &$1^{--}$&0&6631&&$2^{++}$&0&6921\\
       &$1^{+-}$&1&6271&&$0^{-+}$&1&6628&&$1^{+-}$&1&6909\\
       &$2^{++}$&2&6367&&$1^{-+}$&1&6634&&$2^{+-}$&1&6920\\
        2S&&&&&$2^{-+}$&1&6644&&$3^{+-}$&1&6932\\
        &$0^{++}$&0&6782 &&$1^{--}$&2&6635&&$0^{++}$&2&6899\\
      &$1^{+-}$&1&6816& &$2^{--}$&2&6648&&$1^{++}$&2&6904\\
       &$2^{++}$&2&6868& &$3^{--}$&2&6664&&$2^{++}$&2&6915\\
   3S&&&&2P&&&&&$3^{++}$&2&6929\\
        &$0^{++}$&0&7259&&$1^{--}$&0&7091&&$4^{++}$&2&6945\\
       &$1^{+-}$&1&7287 &&$0^{-+}$&1&7100\\
       &$2^{++}$&2&7333&&$1^{-+}$&1&7099\\
      &&&&&$2^{-+}$&1&7098\\
       &&& &&$1^{--}$&2&7113\\
       &&&&&$2^{--}$&2&7113\\
       &&&&&$3^{--}$&2&7112\\
\end{tabular}\end{ruledtabular}
\end{table}

In Table~\ref{ccccexc} we give our predictions for the masses of the
ground and excited states of $cc\bar c\bar c$ tetraquarks
\cite{fgsexc}.  The mass and width of the $X(6900)$ resonance  in
di-$J/\psi$ mass spectrum reported in Ref.~\cite{Aaij:2020fnh} are
\begin{eqnarray*}
  \label{eq:X6900}
  &M[X(6900)]=6905\pm11\pm7~{\rm MeV}, \quad
     \Gamma[X(6900)]=80\pm19\pm33~{\rm MeV} &({\rm Model~1})\cr
      &M[X(6900)]=6886\pm11\pm11~{\rm MeV}, \quad
     \Gamma[X(6900)]=168\pm33\pm69~{\rm MeV}  &({\rm Model~2}),
\end{eqnarray*}
where the difference is based on the treatment of nonresonant
background (see Figure~3 in Reference~\cite{Aaij:2020fnh}). The Model~1 assumes no interference with nonresonant
single-parton scattering (NRSPS), while the Model~2 assumes that the NRSPS
continuum interferes with the broad structure close to the di-$J/\psi$
mass threshold.  We find that this state can be well described either
as the first radial excitation ($2S$) with $J^{PC}=2^{++}$ and the predicted mass
6868~MeV, or as the second orbital excitations ($1D$) $0^{++}$ with
the mass 6899~MeV or $1^{++}$ with mass the 6904~MeV or $2^{++}$ with the mass 6915~MeV. In the
invariant mass spectrum of weighted di-$J/\psi$ candidates
\cite{Aaij:2020fnh}
there is also a hint of another structure around
7.2~GeV. It can correspond to the second radial ($3S$) excitation
$0^{++}$ or/and $2^{++}$ with the predicted masses 7259~MeV and
7333~MeV, respectively.  


\begin{table}
 \caption{Comparison of theoretical predictions for the masses of the
   ground states
 of the neutral  $(QQ)(\bar Q \bar Q)$ tetraquarks (in MeV).}
 \label{tab:cm1}\begin{ruledtabular}
 \begin{tabular}{@{} c@{ } cccccc@{}}
Ref. & \multicolumn{3}{c}{$cc \bar c \bar c$}& \multicolumn{3}{c}{$bb \bar b \bar b$}\\
\cline{2-4} \cline{5-7} & $0^{++}$ & $1^{+-}$ & $2^{++}$ & $0^{++}$ & $1^{+-}$ & $2^{++}$\\
\hline
 \centering{our\cite{fgs}} & 6190 & 6271 & 6367 & 19314 & 19320 & 19330 \\
 \centering{\cite{2004}} & 6477 & 6528 & 6573  &  &  &  \\
 \centering{\cite{2006}} & $6077\pm39$ & $6139\pm38$ & $6194\pm22$  &  &  &  \\
 \centering{\cite{2011}} & 5970 & 6050 & 6220  &  &  &  \\
 \centering{\cite{2012}} & $5300\pm500$ &  &   &  &  &  \\
 \centering{\cite{D12}} & 5966 & 6051 & 6223  & 18754 & 18808 & 18916 \\
 \centering{\cite{SumR2}} & $5990 \pm 80$ & $6050 \pm 80$ & $6090 \pm 80$   & $18840 \pm 90$ & $18840 \pm 90$ & $18850 \pm 90$\\
 \centering{\cite{SumR1}} & $6465\pm5$  & $6440\pm70$ & $6440\pm70$ & $18475\pm15$ & $18430\pm110$ & $18425\pm105$ \\
 \centering{\cite{FullHeavy2018}} &$ <6140$ & &  & 18750 & & \\
 \centering{\cite{lattnrqcd}} & & &  & $>18798$ &$ >19039$ & $>19280$ \\
 \centering{\cite{2018-2}} & & &  & 18800 & & \\
 \centering{\cite{Chiral}} & 6797 & 6899 & 6956 & 20155 & 20212 & 20243\\
 \centering{\cite{E19}} & 5969 & 6021 & 6115 &&&\\ 
 \centering{\cite{WLZ}} &6425&6425&6432&19247 &19247  &19249 \\
 \centering{\cite{FullHeavy2019}} & 6487 & 6500 & 6524 & 19322 & 19329 & 19341 \\
 \centering{\cite{2019-1}} &  &  &  & $18690\pm30$ &  &  \\
 \centering{\cite{Chen}} &&&&19178&19226&19236\\
 \centering{\cite{FullHeavy2019sec}} & 5883 & 6120 & 6246 & 18748 & 18828 & 18900\\
 \centering{\cite{Karliner:2020dta}} & $6192 \pm 25$ & &$6429\pm25$ & $18826 \pm 25$ & & $18956\pm25$\\
 \centering{\cite{Jin:2020jfc}} &6314&6375&6407&19237&19264&19279\\
 \centering{\cite{Lu:2020cns}} &6542&6515&6543&19255&19251&19262\\
 \centering{\cite{DCP}} &6407&6463&6486&19329&19373&19387\\
  \end{tabular}\end{ruledtabular}
\end{table}

In Table~\ref{tab:cm1} we compare our predictions for
the masses of the
   ground states of $QQ\bar Q\bar Q$ tetraquarks with the results of
previous calculations
\cite{2004,2006,2011,2012,lattnrqcd,2018-2,2019-1,D12,SumR1,Karliner:2020dta,SumR2,Chiral,FullHeavy2018,E19,FullHeavy2019,FullHeavy2019sec,
  WLZ,DCP,Chen,Jin:2020jfc,Lu:2020cns}. Our calculation shows that the account of
the diquark structure (size)  weakens the Coulomb-like one-gluon exchange potential, thus increasing tetraquark masses and
reducing spin-spin splittings. We can see from
Table~\ref{tab:cm1} that there are significant
disagreements between different theoretical approaches. Indeed,
Refs.~\cite{2011,2012,lattnrqcd,2018-2,2019-1,D12,SumR2,E19,FullHeavy2019sec} predict heavy
tetraquark masses below or slightly above the thresholds of the decays
to two quarkonia and, thus, stable or significantly suppressed against
fall-apart decays with a very narrow decay width. On the other hand
our model and other approaches predict such tetraquark masses
significantly above these thresholds and, thus, they can be observed only as broad
resonances.

\section{Conclusions}

The calculation of masses of the tetraquarks with
heavy  quarks is reviewed. All considerations are performed in the
framework of the relativistic quark model based on the quasipotential
approach, QCD and  the diquark-antidiquark picture. The dynamical approach
is used, where both diquark and tetraquark masses and wave functions
are obtained by the numerical solution of the quasipotential equation with
the corresponding relativistic  quasipotentials. The structure of the
quark-quark and diquark-antidiquark interactions was fixed from the
previous considerations of meson and baryon properties.
Contrary to most of the considerations, available in the literature,
the diquark is not assumed to be a point-like object. Instead, its
size is explicitly taken into account with the help of the diquark-gluon form factor which is
calculated  as the overlap integral of the diquark wave
functions.  Such a form factor significantly weakens the short-range
Coulomb-like part of the Cornell potential, thus increasing the masses of the tetraquarks and reducing spin splittings. This effect is
especially pronounced for the $bb\bar b \bar b$ tetraquarks since they
have a larger Coulomb contribution due to their  smaller size. 
Note that the approaches with a
point-like diquark substantially underestimate the mass of the doubly
charmed baryon $\Xi_{cc}$, while our model correctly predicted its
mass \cite{efgm} long before its experimental discovery. It is important to pint out that no free adjustable
parameters are introduced. All values of the model parameters 
are kept fixed from the previous calculations of meson and baryon
spectra and decays. This fact significantly improves reliability of
the predictions of our model.

A detailed comparison of our predictions with the current experimental
data was performed. 
It was found that masses of
$X(3872)$, $Z_c(3900)$, $X(3940)$, $Z_{cs}(3985)$, $X(4140)$, $Y(4230)$, $Z_c(4240)$, $Z_c(4250)$,
$Y(4260)$, $Y(4360)$, $Z_c(4430)$, $X(4500)$, $Y(4660)$, $X(4700)$,
$X(4740)$ are
compatible with the masses of hidden-charm tetraquark states with
corresponding quantum numbers. Note that most of these states were
observed after our predictions.
The ground states of
tetraquarks with hidden bottom are predicted to have masses below the 
open bottom threshold and thus should be narrow.  We do not have tetraquark
candidates for charged $Z_b(10610)$ and $Z_b(10650)$, which are
probably molecular states.
Predictions for the masses of bottom counterparts to
the charm tetraquark candidates are given. The experimental search for
these states is an important test of the diquark-antidiquark picture
of heavy tetraquarks.

 In the explicitly exotic $QQ\bar q\bar q$ quark sector the following
 results were obtained \cite{efgl}. All the $(cc)(\bar q\bar
q')$ tetraquarks are predicted to be above the decay threshold into the open
charm mesons. Only  the $I(J^P)=0(1^+)$  state
of $(bb)(\bar u\bar d)$ is 
found to lie below the $BB^*$ threshold. Some of the ground states of
these  tetraquarks are found to have masses just a few tens of MeV above
the thresholds. Thus they, in principle, could be observed as
resonances. 

It was found that the predicted masses of all ground-state $QQ\bar Q\bar Q$
tetraquarks are above the thresholds for decays into two heavy ($Q\bar Q$)
mesons. Therefore they should rapidly fall apart into the two lowest
allowed quarkonium states. Such decays proceed through quark
rearrangements and are not suppressed dynamically or
kinematically. These states should be broad  and are thus difficult to be
observed experimentally. The $2^{++}$
$cc\bar c\bar c$ state with the predicted mass 6367 MeV can correspond
to the broad structure  recently observed by the LHCb Collaboration
\cite{Aaij:2020fnh} in the mass spectrum of  $J/\psi$-pairs produced
in proton-proton collisions.  On the other hand all ground-state
$bb\bar b\bar b$ tetraquarks have masses significantly (400--500 MeV)
higher than corresponding thresholds, and thus should be very
broad. This agrees well with the absence of the narrow beautiful
tetraquarks in the $\Upsilon$-pair production reported by the LHCb \cite{Aaij:2018zrb} and CMS
\cite{Sirunyan:2020txn} Collaborations.

The masses of excited $cc\bar c\bar c$ tetraquarks were
calculated. Lowest radial and orbital excitations between diquark and
antidiquark were considered. It is concluded that the narrow
structure,  $X(6900)$, observed very recently in the $J/\psi$-pair
invariant mass spectrum \cite{Aaij:2020fnh} could be either the first
radial  ($2S$) excitation  or the second orbital ($1D$) excitation of
the $cc\bar c\bar c$ tetraquark, while the structure around 7.2~GeV
could correspond to its second radial ($3S$) excitation.

{  {\bf {Note added:}
} After the manuscript was submitted
  for publication, the~LHCb Collaboration  reported the observation of new
  resonances decaying to $J/\psi K^+$ and $J/\psi \phi$
  \cite{lhcbzcs}. The~charged charmonium-like state with the open
  strangeness $Z_{cs}(4000)^+$ decaying to $J/\psi K^+$, with~the mass
  $4003\pm6^{+4}_{-14}$ MeV and spin-parity $1^+$, was observed with
  large significance. Although~its mass is consistent with the mass of
  the $Z_{cs}(3985)^-$ state previously observed by the BESIII
  experiment in the $D^-_s D^{*0}+D^{*-}_s D^0$ mass distribution
~\cite{Zcs}, the~decay width measured by the LHCb is about an order
  of magnitude broader than the width reported by the
  BESIII. Therefore, the~LHCb concludes~\cite{lhcbzcs} that there is
  no evidence that the $Z_{cs}(4000)^+$ state is the same as the
  $Z_{cs}(3985)^-$ state. This conclusion is in accordance with our
  predictions~\cite{efgtetr,efgextetr}. Indeed, both states can be
  naturally interpreted in our model as the $1^+$ state composed from
  scalar and axial vector diquarks with the predicted mass 3982 MeV
  for the $Z_{cs}(3985)^-$ and as the $1^+$ state composed from  axial
  vector diquark and antidiquark with the predicted mass 4004 MeV for
  the $Z_{cs}(4000)^+$ state (see Table~\ref{tab:ecmass}). The~new
  $1^+$ $X(4685)$ state decaying to $J/\psi \phi$ final state with the
  measured mass $4684\pm7^{+13}_{-16}$~\cite{lhcbzcs} agrees with our
  prediction for the $2S$ $cs\bar c\bar s$ state, composed from scalar
  and axial vector diquarks, with~the mass 4665 MeV. The~new $X(4630)$
  state~\cite{lhcbzcs} with the measured mass $4626\pm16^{+18}_{-110}$
  and quantum numbers $1^-$ or $2^-$, where the first assignment is preferred
  at a $3\sigma$ level, could be interpreted in our model as the $1P$
  state $1^-$, composed from scalar and axial vector diquarks, with~the
  predicted mass 4514 MeV, or~as the $2^-$ state, composed from axial
  vector diquark and antidiquark, with~the predicted mass 4598
  MeV. The~broad $Z_{cs}(4220)^+$ state with the measured mass
  $4216\pm24^{+43}_{-30}$ MeV with quantum numbers $1^+$ or $1^-$
  (with a 2$\sigma$ difference in favor of the first hypothesis)
  \cite{lhcbzcs} is consistent with our prediction for the $1P$ state
  $1^-$, composed from the scalar diquark and antidiquark, and~the mass 4350 MeV.}

\acknowledgments{The authors are grateful to  D. Ebert for very
  fruitful and pleasant collaboration in developing the diquark-antidiquark model
  of tetraquarks. We express our gratitude to
  A. Berezhnoy and M. Wagner for 
  valuable  discussions. We thank the organizers of the XXXII
  International Workshop on High Energy Physics "Hot problems of
  Strong Interactions”, especially  V. Petrov
  and R. Zhokhov,  for the
invitation to participate in such an interesting and productive meeting.}

\end{document}